\documentclass[aps,pra,twocolumn,floatfix,superscriptaddress,noeprints]{revtex4-1}
\usepackage[utf8]{inputenc}
\usepackage{amsmath,amssymb,amsfonts,bbm,graphicx,hyperref,times,color}
\def\Tr{\hbox{Tr}}
\def\sigmaCM{\boldsymbol{\sigma}}

\begin{document}
\title{Ultimate limits for quantum magnetometry via time-continuous measurements}
\author{Francesco Albarelli}
\email{francesco.albarelli@unimi.it}
\affiliation{Quantum Technology Lab, Dipartimento di Fisica, Universit\`a degli Studi di Milano, 20133 Milano, Italy}
\author{Matteo A. C. Rossi}
\email{matteo.rossi@unimi.it}
\affiliation{Quantum Technology Lab, Dipartimento di Fisica, Universit\`a degli Studi di Milano, 20133 Milano, Italy}
\author{Matteo G. A. Paris}
\email{matteo.paris@fisica.unimi.it}
\affiliation{Quantum Technology Lab, Dipartimento di Fisica, Universit\`a degli Studi di Milano, 20133 Milano, Italy}
\affiliation{INFN, Sezione di Milano, I-20133 Milano, Italy}
\author{Marco G. Genoni}
\email{marco.genoni@fisica.unimi.it}
\affiliation{Quantum Technology Lab, Dipartimento di Fisica, Universit\`a degli Studi di Milano, 20133 Milano, Italy}

\begin{abstract}
We address the estimation of the magnetic field $B$ acting
on an ensemble of atoms with total spin $J$ subjected to collective transverse
noise. By preparing an initial spin coherent state,
for any measurement performed after the evolution, the mean-square
error of the estimate is known to scale as $1/J$, i.e. no quantum
enhancement is obtained. Here, we consider the possibility of
continuously monitoring the atomic environment, and conclusively show that
strategies based on time-continuous non-demolition
measurements followed by a final strong measurement may achieve
Heisenberg-limited scaling $1/J^2$ and also a monitoring-enhanced
scaling in terms of the interrogation time. We also find
that time-continuous schemes are robust against detection losses,
as we prove that the quantum enhancement can be recovered also
for finite measurement efficiency.
Finally, we analytically prove the optimality
of our strategy.
\end{abstract}

\maketitle

\section{Introduction}\label{s:intro}
Recent developments in the field of quantum metrology have shown how
quantum probes and quantum measurements allow one to achieve parameter
estimation with precision beyond that obtainable by any classical scheme~\cite{GiovannettiNatPhot,RafalPO}. The estimation of the strength of a
magnetic field is a paradigmatic example in this respect, as it can
be mapped to the problem of estimating the Larmor frequency for an
atomic spin ensemble~\cite{Wasilewski2010,Koschorreck2010,Sewell2012,Ockeloen2013,Sheng2013,
Lucivero2014,Muessel2014}.
\par
As a matter of fact, if the system is initially prepared in a
spin coherent state, the mean-squared error of the
field estimate scales, in terms of the total spin number $J$, as $1/J$,
which is usually referred to as the standard quantum limit (SQL) to precision.
If quantum resources, such as spin squeezing or entanglement
between the atoms of the spin ensemble, are exploited, one observes a quadratic enhancement and achieves the so-called {\em Heisenberg scaling}, i.e. $1/J^2$~\cite{Wineland92,Bollinger96}.
On the other hand, it has been proved that
such ultimate quantum limit may be easily lost in the presence of noise~\cite{Huelga97} and that typically a SQL-like scaling is observed, with
the quantum enhancement reduced to a constant factor.
These observations have been rigorously translated into a set of no-go theorems~\cite{EscherNatPhys,KolodynskyNatComm},
which fostered several attempts to circumvent them. In particular, it has been shown
how one can restore a super-classical scaling in the context of
frequency estimation, for specific noisy evolution and/or by optimizing
the strategy over the interrogation time~\cite{Matsuzaki2011,Chin12,Chaves13,Brask15,Smirne16}, or
by exploiting techniques borrowed from the field of quantum
error-correction~\cite{Kessler14,Dur14,Arrad14}.
\par
In this manuscript, we put forward an alternative approach:
we assume to start the dynamics with a \emph{classical} state
that is monitored continuously in time via the interacting environment
\cite{WisemanMilburn,SteckJacobs}. The goal is to recover
 the information on the parameter leaking into the environment
 and simultaneously to exploit the back action of the measurement to
 drive the system into more sensitive conditional states
\cite{WisemanSqueezing,Thomsen2002,WisemanMancini,SerafozziMancini,DohertyPRL2011,Tempura,Ravotto,Hofer2015,Levante,Calypso}.
This approach has received much attention recently
\cite{GammelmarkCRB,GammelmarkQCRB,KiilerichPC,KiilerichHomodyne,Catana2014,Gefen2016,Plenio2016,Cortez2017}
also in the context of quantum magnetometry \cite{Geremia2003,Stockton2004,Auzinsh2004,Molmer2004,Madsen2004,Chase2009}.

\begin{figure}[h!]
\centering
\includegraphics[width=0.7\columnwidth]{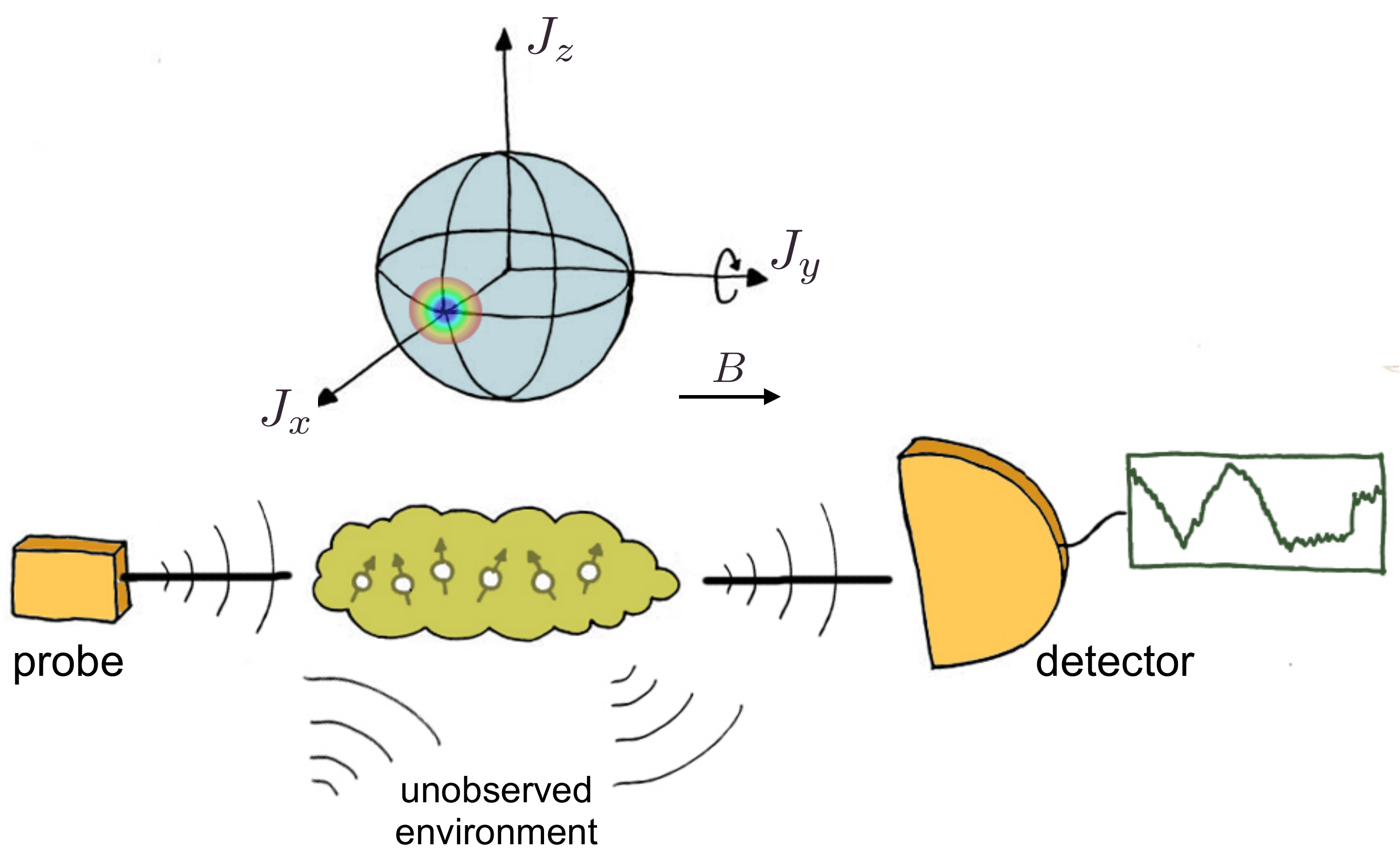}
\caption{{\bf Atomic magnetometry via time-continuous
measurements} - an atomic ensemble, prepared in a spin-coherent state
aligned to the $x$-direction and placed in a constant magnetic field $B$ pointing
in the $y$-direction, is coupled to train of probing fields that are
continuously monitored after the interaction with the sample.}
\label{fig:figure_magnetometry}
\end{figure}

Here we rigorously address the performance of these protocols, depicted in Fig.~\ref{fig:figure_magnetometry},
taking into account the information obtained via the time-continuous non-demolition measurements on the environment,
as well as the information obtainable via a strong (destructive)
measurement on the conditional state of the system.
In particular, in the limit of large spin, we derive an analytical formula for the ultimate bound on the
mean-squared error of any unbiased estimator, and conclusively show that, for
experimentally relevant values of the dynamical parameters,
one can observe a Heisenberg-like scaling.\\
\par
Remarkably, at variance with most of the protocols proposed for quantum
magnetometry, and in general for frequency estimation, one does not need
to prepare an initial spin-squeezed state. The Heisenberg scaling is in
fact obtained also for an initial {\em classical} spin coherent state,
thanks to the spin squeezing generated by time-continuous measurements'
back-action. Finally, we analytically prove that the ultimate quantum
limit for noisy magnetometry in the presence of collective transverse noise
\cite{GammelmarkQCRB} is in fact saturated by our strategy, i.e.
one does not need to implement more involved strategies, e.g.
jointly measuring the conditional state of the system and the
output modes of the environment at different times.
\par
The paper is organized as follows. In Sec.~\ref{s:tcqcrb}, we present the
quantum Cram\'er-Rao bounds that hold for noisy metrology, with emphasis
on estimation strategies based on time-continuous,
non-demolition measurements and a final \emph{strong} measurement on
the corresponding conditional quantum states. In Sec.~\ref{s:magneto},
we introduce the physical setting for the estimation of a magnetic
field via a continuously monitored atomic ensemble.
In particular, we focus on the case of large total spin, where a
Gaussian picture is able to describe the whole dynamics. In Sec.~\ref{s:results}, we present the main results: we first calculate the
classical Fisher information corresponding to the photoccurent obtained
via the time-continuous monitoring of the environment, and we discuss
how to attain the corresponding bound via Bayesian estimation. We then
address the possibility of performing also a strong measurement on the
conditional state of the atomic ensemble, and derive the ultimate limit
on this kind of estimation strategy, quantified by an effective quantum Fisher information. Upon studying this quantity, we observe how, in the relevant parameters' regime, the Heisenberg limit can be effectively restored, also discussing
the effects of non-unit monitoring efficiency, corresponding to the loss
of photons before the detection. Finally, we also prove the optimality
of our measurement strategy in the case of ideal detectors.
Section~\ref{s:conclusion} closes the paper with some concluding
remarks.
\section{Quantum Cramér-Rao bounds for time-continuous homodyne monitoring} \label{s:tcqcrb}
A classical estimation problem consists in inferring the value of a parameter
$\theta$ from a number $M$ of measurement outcomes $\chi =
\{x_1,x_2,\dots,x_M\}$ and their conditional distribution
$p(x|\theta)$. We define an estimator $\hat{\theta}(\chi)$ a function from the measurement outcomes to the possible values of $\theta$ and we dub it asymptotically unbiased when, in the limit of large number of repetitions of the experiment $M$, its average is equal to the true value, i.e. $\int d\chi\, p(\chi|\lambda) \hat{\theta}(\chi) =\theta$, where $p(\chi |\lambda) = \Pi_{j=1}^M p(x_j | \theta)$. The Cram\'er-Rao theorem states that the variance of any unbiased estimator is lower bounded as ${\rm Var}_{\hat{\theta}}(\theta) = \left( M \mathcal{F}[p(x|\theta)] \right)^{-1} $, where
\begin{align}
\mathcal{F}[p(x|\theta)] = \int dx\, p(x|\lambda) (\partial_\theta \log p(x|\lambda) )^2
\end{align}
denotes the classical Fisher information (FI).
\par
In the quantum realm, the conditional probability distribution reads
$p(x|\theta) = \Tr[\varrho_\theta \Pi_x]$, where $\varrho_\theta$ is
the quantum state of the system labeled by the parameter $\theta$,
and $\Pi_x$ is a POVM operator describing the quantum measurement.
One can prove that the FI corresponding to any POVM is
upper bounded $\mathcal{F}[p(x|\theta)]\leq \mathcal{Q}[\varrho_\theta]$,
where $\mathcal{Q}[\varrho_\theta] = \Tr[\varrho_\theta L_\theta^2]$ is the
quantum Fisher information (QFI), and $L_\theta$ is the so-called
symmetric logarithmic derivative, which can be obtained by solving the
equation $2 \partial_\theta \varrho_\theta = L_\theta \varrho_\theta
+ \varrho_\theta L_\theta$ \cite{HelstromBook,CavesBraunstein,MatteoIJQI}.
The QFI depends on the quantum state $\varrho_\theta$ only, and thus
poses the ultimate bound on the precision of the estimation of $\theta$.
Moreover, in the single parameter case the bound is always achievable,
that is, there exists a (projective) POVM such that the corresponding
classical FI equals the QFI.\\
\par

%
In this manuscript we consider a quantum system evolving according to a given Hamiltonian $\hat{H}_\theta$ characterised by the parameter we want to estimate, and coupled to a bosonic environment at zero temperature described by a {\em train} of input operators $\hat{a}_{in}(t)$, satisfying the commutation relation
$[\hat{a}_{in}(t),\hat{a}_{in}^\dag(t')]=\delta(t-t')$, via an interaction Hamiltonian $\hat{H}_{int}(t) =\hat{c} \hat{a}_{in}^{\dag}(t) + \hat{c}^\dag \hat{a}_{in}(t)$ ($\hat{c}$ being a generic operator in the system Hilbert space) \cite{WisemanMilburn}. By tracing out the environment, the unconditional dynamics of the system is described by the Lindblad master equation
\begin{align}
\frac{d\varrho}{dt} = \mathcal{L}\varrho = - i [\hat{H}_{\theta} , \varrho] +  \mathcal{D}[\hat{c}] \varrho \:, \label{eq:ME}
\end{align}
where $\mathcal{D}[c]\varrho = \hat{c} \varrho \hat{c}^{\dag} - (\hat{c}^\dag \hat{c} \varrho + \varrho \hat{c}^\dag \hat{c} )/2$.\\
If one performs a homodyne detection of a quadrature $\hat{x}_{out}(t) = \hat{a}_{out}(t) + \hat{a}_{out}^\dag(t) $ 
on the output operators, {\em i.e.} on the environment just after the interaction with the system, one obtains that the dynamics of the system quantum state $\varrho^{(c)}$ conditioned on the measurement results (we will omit the dependence of the measured photocurrent ${\bf y}_t$), is described by the stochastic master equation \cite{WisemanMilburn}
\begin{align}
d\varrho^{(c)} = - i [\hat{H}_{\theta} , \varrho^{(c)}]\,dt +  \mathcal{D}[\hat{c}] \varrho^{(c)} \,dt + \sqrt{\eta} \mathcal{H}[\hat{c}]\varrho^{(c)} \, dw_t \:. \label{eq:SME}
\end{align}
Here $\eta$ denotes the efficiency of the detection, $dw_t$ is a stochastic Wiener increment (s.t. $dw_t^2=dt$), and $\mathcal{H}[c]\varrho^{(c)} = \hat{c} \varrho^{(c)} + \varrho^{(c)} \hat{c}^\dag - \Tr[\varrho^{(c)} (\hat{c} +\hat{c}^\dag)]\varrho^{(c)}$ (notice that in principle one could consider other measurement strategies different from homodyne, yielding a different superoperator).
The corresponding measurement record during a time step $t\rightarrow t+dt$ is given by the infinitesimal current
\begin{align}
dy_t= \sqrt{\eta}\, \Tr[\varrho^{(c)} (\hat{c}+\hat{c}^\dag)] \, dt + dw_t. \label{eq:dyt}
\end{align}
With the help of such measurement strategies, one can estimate the value of the parameter $\theta$ both from the measured photocurrent  ${\bf y}_T = \int_0^T dy_t$, and from a final {\em strong} (destructive) measurement on the conditional state $\varrho^{(c)}$. In this case, as we explicitly show in~\ref{appendix0} (in general both for the classical and quantum case), the proper quantum Cram\'er-Rao bound reads
\begin{align}
{\rm Var}_{\hat{\theta}}(\theta) \geq \frac{1}{M\left( \mathcal{F}[p({\bf y}_T)] + \mathbbm{E}_{p({\bf y}_T)}\left[ \mathcal{Q}[\varrho^{(c)}]  \right]\right) }\, , \label{eq:tcqcrb}
\end{align}
where the first term at the denominator $\mathcal{F}[p({\bf y}_T)]$ is the FI
corresponding to the classical photocurrent ${\bf y}_T$, while
the second term is the average of the QFI
for the conditional state $\mathcal{Q}[\varrho^{(c)}]$
over all the possible trajectories, i.e. on all the possible measurement outcomes for the photocurrent.\\
The classical FI $\mathcal{F}[p({\bf y}_T)]$  can be calculated as described in \cite{GammelmarkCRB} by evaluating
\begin{align}
\mathcal{F}[p({\bf y}_T)]= \mathbbm{E}_{p({\bf y}_T)} [ \Tr[\tau]^2 ] \, ,
\end{align}
where the operator $\tau$ evolves according to the stochastic master equation
\begin{align}
d\tau &= - i [\hat{H}_{\theta}, \tau] \, dt - i [(\partial_\theta \hat{H}_{\theta}), \varrho] \, dt + 
\\ & +\mathcal{D}[\hat{c}] \tau \, dt + ( \hat{c}\tau + \tau \hat{c}^\dag ) dw_t \:.
\end{align}
The conditional states $\varrho^{(c)}$ at time $T$ can be obtained by integrating~\eqref{eq:SME}, for a certain stream of outcomes $\mathbf{y}_T$; then one can first calculate the corresponding quantum Fisher information $\mathcal{Q}[\varrho^{(c)}]$, and, numerically or when possible analytically, its average over all the possible trajectories explored by the quantum system due to the homodyne monitoring.

A more fundamental quantum Cramér-Rao bound that applies in this physical setting has been derived in \cite{GammelmarkQCRB}, by considering the QFI obtained from the unitary dynamics of the global pure state of system and environment.
This QFI is obtained by optimizing over all possible POVMs, i.e. one also considers the possibility of performing non-separable (entangled) measurements over the system and all the output modes $\hat{a}_{out}(t)$ at different times.
On the other hand, in the previous setting the estimation strategies were restricted to the more experimentally friendly case of sequential/separable measurements on the output modes and on the final conditional state of the system.
\par
The QFI expressing this ultimate QCRB is by definition
\begin{equation}
\label{eq:ultimQFI}
	\mathcal{\overline{Q}}_{\mathcal{L}}(\theta) = 4 \partial_{\theta_1} \partial_{\theta_2} \log \left( \left| \langle \psi(\theta_1) | \psi(\theta_2) \rangle \right| \right) \big|_{\theta_1 = \theta_2 = \theta},
\end{equation}
where $\langle \psi(\theta_1) | \psi(\theta_2) \rangle$ is the fidelity between the \emph{global} state of system and environment for two different values of the parameter, and where we have highlighted its dependence on the superoperator $\mathcal{L}$ that defines the unconditional master equation~\eqref{eq:ME}. The key insight is that this fidelity can be determined by using operators acting on the system only \cite{GammelmarkQCRB,Macieszczak2016} and it can be expressed as the trace of an operator $\Tr\left[ \bar{\rho} \right]=\langle \psi(\theta_1) | \psi(\theta_2) \rangle$, which obeys the following generalized master equation
\begin{equation}
	\label{eq:genME}
	\frac{d \bar{\rho}}{d t} = -i \left( \hat{H}_{\theta_1} \bar{\rho} - \bar{\rho} \hat{H}_{\theta_2} \right) + \mathcal{D}\left[ \hat{c} \right] \bar{\rho} \,.
\end{equation}
As before, we already assumed that the dependence on the parameter lies only in the system Hamiltonian $\hat{H}_\theta$ and that we have a single jump operator $\hat{c}$. We remark that the operator $\bar{\rho}$ is not a proper density operator representing a quantum state, except in the limit case $\theta_1 \to \theta_2$, where we recover the standard master equation~\eqref{eq:ME}.
\section{Quantum magnetometry: the physical setting} \label{s:magneto}
We address the estimation of the intensity of a static and constant magnetic field $B$ acting on a ensemble of $N$ two-level atoms that are continuously monitored \cite{Geremia2003,Stockton2004,Auzinsh2004,Molmer2004}, as depicted in Fig.~\ref{fig:figure_magnetometry}. The atomic ensemble can be described  as a system with total spin $J=N/2$ with collective spin operators defined as
$\hat{J}_{\alpha} = \frac{1}{2} \sum_{i=0}^N \sigma_{i \alpha}$, where $\alpha = x,y,z$
and $\sigma_{i\alpha}$ denotes the Pauli matrices acting on the $i$-th spin.
The collective operators obey the same angular momentum commutation rules $[ \hat{J}_i, \hat{J}_j ] = i \varepsilon_{ijk} \hat{J}_k$, where $\varepsilon_{ijk}$ is the Levi-Civita symbol. We remark that in the present manuscript we choose units such that $\hbar = 1$.

We assume that the atomic sample is coupled to a electromagnetic mode $a_{in}(t)$ corresponding either to a cavity mode in a strongly driven and heavily damped cavity \cite{Thomsen2002}, or analogously to a far-detuned traveling mode passing through the ensemble \cite{Molmer2004}. By considering an interaction Hamiltonian $\hat{H}_{int} = \sqrt{\kappa} \hat{J}_z ( \hat{a}_{in}(t) + \hat{a}_{in}^\dag(t) )$ and if these {\em environmental} light modes are left unmeasured, the evolution of the system is expressed by \eqref{eq:ME}, which in this case corresponds to a collective transverse noise on the atomic sample,
\begin{align}
\frac{d \varrho}{d t} &= \mathcal{L}_{tn} \varrho = - i \gamma B [ \hat{J}_y, \varrho ] + \kappa \mathcal{D} [ \hat{J}_z ] \varrho,
\label{eq:ME_B}
\end{align}
where the constants $\kappa$ and $\gamma$ represent respectively the strength of the coupling with the noise and with the magnetic field, that is directed on the $y$-axis and thus perpendicular to the noise generator. At $t=0$ we consider the system prepared in a spin coherent state, i.e. a tensor product of single spin states (qubits) directed in the positive $x$ direction,
\begin{equation}
|\psi(0) \rangle = \bigotimes_{k=0}^{N} | + \rangle_k = | J , J\rangle_x,
\end{equation}
where $|+\rangle$ is the eigenstate of $\sigma_x$ with eigenvalue $+1$. We thus have that the spin component on the $x$ direction attains the macroscopic value $\langle \hat{J}_x (0)\rangle=J$.
The unconditional dynamics of $\langle \hat{J}_x \rangle$ is obtained by applying the operator $\hat{J}_x$ to both sides of Eq.~\eqref{eq:ME_B} and then taking the trace.
The result is the following equation describing damped oscillations
\begin{align}
\frac{d \langle {\hat{J}_x (t)} \rangle}{dt} &=  \gamma B \langle \hat{J}_z (t) \rangle - \frac{\kappa}{2} \langle  \,\hat{J}_x (t) \rangle \,,
\end{align}
where we observe how the the dissipative and unitary parts of the dynamics are respectively shrinking the spin vector $\langle \vec{\hat{J}} \rangle$ and causing its Larmor precession around the $y$-axis. In the following we will assume to measure {\em small} magnetic fields, such that $\gamma B t \ll 1$ and we can approximate the solution of the previous equation as
\begin{equation}
\label{eq:dampedJx}
\langle \hat{J}_x (t) \rangle \approx \langle \hat{J}_x (0) \rangle e^{- \kappa t / 2} = J e^{- \kappa t / 2}  \,.
\end{equation}
If the light modes are continuously monitored via homodyne measurements at the appropriate phase, one allows a continuous ``weak'' measurement of $\hat{J}_z$; the corresponding stochastic master equation~\eqref{eq:SME} for finite monitoring efficiency $\eta$ reads
\begin{equation}
\label{eq:stochME_spin}
\begin{split}
d \varrho^{(c)}  =& - i \gamma B [ \hat{J}_y, \varrho^{(c)} ] \mathrm{d} t + \kappa \mathcal{D} [ \hat{J}_z ] \varrho^{(c)} d t  + \sqrt{\eta \kappa} \mathcal{H} [ \hat{J}_z ] \varrho^{(c)} d w_t\, ,
\end{split}
\end{equation}
while the measurement result at time $t$ corresponds to an infinitesimal photocurrent $dy_t = 2\sqrt{\eta\kappa}\, \Tr[\varrho^{(c)} \hat{J}_z] dt + dw_t$. It is important to remark how the collective noise characterizing the master equation~\eqref{eq:ME_B} describes the dynamics also in experimental situations where no additional coupling to the atomic ensemble, with the purpose of performing continuous monitoring, is engineered \cite{Plankensteiner2016,DallaTorre2013,Dorner2012a}.
In this respect, assuming a non-unit efficiency $\eta$ corresponds to considering both homodyne detectors that are not able to capture all the photons that have interacted with the spin, and environmental degrees of freedom, causing the same kind of noisy dynamics, that cannot be measured during the experiment.
\par
Let us now consider the limit of large spin $J \gg 1$. In this case,
the dynamics may be effectively described with the Gaussian formalism
as long as $\langle \hat{J}_x (t) \rangle \approx J$, i.e. for times $t$ small enough to guarantee that
$ \kappa t \lesssim 1 $.  We define the effective quadrature operators of the atomic sample, satisfying the canonical commutation relation $[\hat{X},\hat{P}]=i$, as~\cite{Madsen2004,Molmer2004}
\begin{equation}
\hat{X} = \hat{J}_y / \sqrt{\bar{J}_t } \qquad
\hat{P} = \hat{J}_z /\sqrt{\bar{J}_t }  \,,
\end{equation}
where $\bar{J}_t \equiv | \langle \hat{J}_x (t) \rangle |$
(notice that in the limit of large spin $J$ we can safely consider the unconditional average value $\langle \hat{J}_x (t) \rangle$, as the stochastic correction obtained via~\eqref{eq:stochME_spin} would be negligible).
In the Gaussian description the initial state $|\psi (0) \rangle$ corresponds to the vacuum state $ (\hat{X} + i \hat{P}) |0\rangle = |0\rangle$, which is Gaussian.
As the stochastic master equation~\eqref{eq:stochME_spin} becomes quadratic in the canonical operators (and thus preserves the Gaussian character of states)
\begin{equation}
\label{eq:stochME_gaus}
\begin{split}
d\varrho^{(c)} &= - i \gamma B \sqrt{\bar{J}_t} \left[ \hat{X}, \varrho^{(c)} \right] \, dt + \\ 
& + \kappa \bar{J}_t \mathcal{D} [ \hat{P} ] \, \varrho^{(c)} dt + \sqrt{\bar{J}_t \eta \kappa } \,\mathcal{H}[ \hat{P} ] \, \varrho^{(c)} dw_t \,,
\end{split}
\end{equation}
the whole dynamics can be equivalently rewritten in terms of first and second moments only \cite{diffusone,WisemanDoherty} (see~\ref{appendix1}
for the equations describing the whole dynamics in the Gaussian picture).
As it will be clear in the following, due to the nature of the coupling, in order to address the estimation of $B$, we only need the behaviour of the mean and the variance of the atomic momentum quadrature $\hat{P}$ calculated on the conditional state $\varrho^{(c)}$, which follows the equations
\begin{align}
&d \langle \hat{P} (t) \rangle_c = - B \gamma \sqrt{J e^{- \frac{\kappa t}{2}} } dt + 2 \mathrm{Var}_c[ \hat{P} (t) ]\sqrt{ \eta \kappa J e^{-\frac{\kappa t}{2}}}   dw_t \,, \label{eq:firstmom}\\
 & \frac{  d\mathrm{Var}_c[ \hat{P} (t) ]}{dt} =	-4 \eta \kappa  J e^{-\frac{\kappa t}{2}} \left( \mathrm{Var}_c[ \hat{P} (t) ] \right)^2. \label{eq:secondmom}
\end{align}
The differential equation for the conditional second moment is deterministic and can be solved analytically. For an initial vacuum state, {\em i.e.} with $\mathrm{Var}[ \hat{P} (0)] = \frac{1}{2} $, we obtain the following solution
\begin{equation}
\label{eq:varsoldamp}
\mathrm{Var}_c[ \hat{P} (t)] = \frac{1}{8 \eta  J \left(1-e^{-\frac{\kappa t}{2}} \right)+2} \,,
\end{equation}
that shows how the conditional state of the atomic sample is deterministically driven by the dynamics into a spin-squeezed state.

\section{Results}\label{s:results}
Here we will present our main results, that is the derivation of ultimate quantum limits on noisy magnetometry via time-continuous measurements of the atomic sample. We will first evaluate the classical Fisher information $\mathcal{F}[{\bf y}_t]$ corresponding to the information obtainable from the photocurrent, and we will also show how the corresponding bound can be achieved via Bayesian estimation. We will then evaluate the second term appearing in the bound, corresponding to the information obtainable via a {\em strong} measurement on the conditional state of the atomic sample. This will allow us to discuss the ultimate limit on the estimation strategy via the effective quantum Fisher information: we will focus on the scaling with the relevant parameters of the experiment, {\em i.e.} with the total spin number $J$ and the monitoring time characterizing each experimental run $t$, and we will address the role of the detector efficiency $\eta$. \\
%
\subsection{Analytical FI corresponding to the time-continuous photocurrent}
As discussed before, the measured photocurrent $\mathbf{y}_t$ obtained via continuous homodyne detection can be used to extract information about the system and to estimate parameters which appear in the dynamics. The ultimate limit on the precision of this estimate is quantified by the FI $\mathcal{F}[p({\bf y}_t)]$.
Given the Gaussian nature and the simple dynamics of the problem we can compute it analytically in closed form, by applying the results of \cite{Genoni2017}. As we describe in more detail in~\ref{appendix1}, one obtains the formula
\begin{equation}
\mathcal{F}[p({\bf y}_t)] =  2 \eta \kappa J e^{-\kappa t/2} \, \mathbbm{E}_{p({\bf y}_t)} \left[ \left( \partial_B \langle \hat{P} (t) \rangle_c \right)^2  \right] \,. \label{eq:tcfisher_magenot}
\end{equation}
By considering \eqref{eq:firstmom} and remembering that $\mathrm{d} w_t = \mathrm{d} y_t - \sqrt{2 \eta \kappa J e^{-\frac{\kappa t}{2}}} \langle \hat{P}(t) \rangle_c \mathrm{d} t $, one obtains that the time evolution of the derivative of the conditional first moment $\langle \hat{P} (t) \rangle_c$ w.r.t. to the parameter $B$, can be written as
\begin{equation}
\label{eq:derR}
\begin{split}
& \frac{d \left( \partial_B \langle \hat{P} (t) \rangle_c \right)}{dt} = 
\\ & =  - \gamma \sqrt{J e^{- \kappa t / 2}} - 4 \mathrm{Var}_c[ \hat{P} (t)]  \eta \kappa J e^{- \kappa t / 2}  \left( \partial_B \langle \hat{P} (t) \rangle_c \right).
\end{split}
\end{equation}
where $\mathrm{Var}_c[ \hat{P} (t)]$ is obtained from Eq.~\eqref{eq:varsoldamp}. We thus observe that the evolution is deterministic and one can easily derive its analytical solution. By applying Eq.~\eqref{eq:tcfisher_magenot}, as the average over the trajectories is not needed, we readily obtain the following analytical formula for the FI
\begin{equation}
\label{eq:fisher_cont}
\begin{split}
&\mathcal{F}[p({\bf y}_t)]=
\frac{64 \gamma ^2 \eta  J^2 e^{ -\kappa t} \left(e^{\frac{\kappa  t}{4}}-1\right)^3}{9 \kappa ^2 \left[(4 \eta  J+1) e^{\frac{\kappa  t}{2}}-4 \eta  J\right]} \cdot \\
& \cdot [-4 \eta  J-12 \eta  J e^{\frac{\kappa  t}{4}}+3 (4 \eta  J+3) e^{\frac{\kappa  t}{2}}+(4 \eta  J+3) e^{\frac{3 \kappa  t}{4}}].
\end{split}
\end{equation}
As intuitively expected, this is a monotonically increasing function of $t$, since the partial derivative is always positive. To get some insight into this expression we first report the leading term for $t\to 0$
\begin{equation}
\label{eq:fisher_cont_expansionT}
\mathcal{F}[p({\bf y}_t)] \approx \frac{4}{3} J^2 \gamma^2 \kappa t^3,
\end{equation}
where we explicitly see both Heisenberg scaling $J^2$ and a monitoring-enhanced time scaling $t^3$. 
We can get further intuition about this expression by expanding it around $J=\infty$, the limit in which the Gaussian approximation becomes exact.
The leading order in this other expansion is quadratic in $J$, thus showing again Heisenberg scaling, irregardless of $t$:
\begin{equation}
\label{eq:fisher_cont_expansionJ}
\mathcal{F}[p({\bf y}_t)] \approx \frac{64 \gamma ^2 \eta  J^2 e^{ - \kappa t} \left(e^{\frac{\kappa  t}{4}}-1\right)^3 \left(4 e^{\frac{\kappa  t}{4}}+e^{\frac{\kappa  t}{2}}+1\right)}{9 \kappa ^2 \left(e^{\frac{\kappa  t}{4}}+1\right)};
\end{equation}
this last approximations actually reproduces the behavior of the function quite well in the range of parameters we will consider in the following.

\begin{figure}[h!]
\centering
\includegraphics[width=.7\columnwidth]{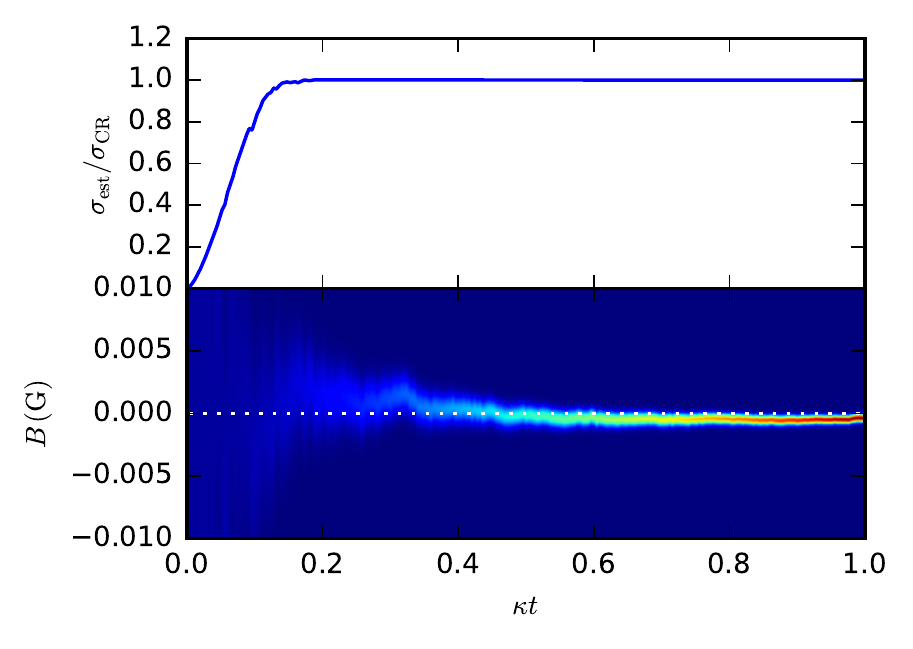}
\caption{{\bf Bayesian estimation of $B$ from a single simulated experiment} - the data shown in the plots are obtained as a function of $\kappa t$, for $\gamma/\kappa=1\, \mathrm{G}^{-1}$, $J=10^4$ and $\eta=1$; the prior distribution of the parameter $B$ is uniform in the interval $[-0.01,0.01] \, \mathrm{G}$, and the true value is $B_{\mathrm{true}}=0\,\mathrm{G}$. In the top panel we show the ratio between the standard deviation of the posterior distribution and the standard deviation predicted by the Cramér-Rao bound. In the bottom panel we show the posterior distribution as a function of time, the constant white dashed line marks the value $B_{\mathrm{true}}$.
}\label{fig:posterior_time}
\end{figure}

We now want to show that one can achieve this classical Cram\'er-Rao bound from the time-continuous measurement outcomes obtained via an appropriate estimator.
In Figure~\ref{fig:posterior_time} we indeed show the posterior distribution as a function of time for a single experimental run, obtained after a Bayesian analysis (see~\ref{appendixBayes} for details). We observe how the distribution gets narrower in time around the true value and we also explicitly show that its standard deviation $\sigma_{\mathrm{est}}$ converges to the one predicted by the Cramér-Rao bound $\sigma_{\mathrm{CR}}(t)=\mathcal{F}[p({\bf y}_t)]^{-1/2}$. In the initial part of the dynamics the values of $\sigma_{\mathrm{est}}$ are smaller than the corresponding
$\sigma_{\mathrm{CR}}$: this is due to the choice of the prior distribution, being narrower than the likelihood and thus implying some initial knowledge on the parameter which is larger than the one obtainable for small monitoring time.
%

\subsection{Quantum Cram\'er-Rao bound for noisy magnetometry via time-continuous measurements}
In order to evaluate the quantum Cram\'er-Rao bound in Eq.~\eqref{eq:tcqcrb} we now need to consider the second term $\mathbbm{E}_{p({\bf y}_T)}\left[ \mathcal{Q}[\varrho^{(c)}]  \right]$, corresponding to the information obtainable via strong quantum measurement on the conditional state of the system. The conditional state $\varrho^{(c)}$ is Gaussian and has a dependence on the parameter $B$ only in the first moments. Therefore the corresponding QFI can be evaluated as prescribed in~\cite{Pinel2013} (see~\ref{appendix1} for more details) obtaining,
\begin{equation}
\mathcal{Q}[\varrho^{(c)}] = \frac{ \left( \partial_B \langle \hat{P}(t) \rangle_c \right)^2 }{\mathrm{Var}_c[ \hat{P} (t)]}.
\label{eq:QFI_formula}
\end{equation}
Since, as we proved before, the evolution of both $\partial_B \langle \hat{P}(t) \rangle_c$ and $\mathrm{Var}_c[ \hat{P} (t)]$ is deterministic, the average over all possible trajectories is also in this case trivial and we have $\mathbbm{E}_{p({\bf y}_T)}\left[ \mathcal{Q}[\varrho^{(c)}]  \right] = \mathcal{Q}[\varrho^{(c)}]$. By exploiting the analytical solution for both quantities, the QFI reads
\begin{equation}
\label{eq:QFI_t}
\mathcal{Q}[\varrho^{(c)}] = \frac{32 \gamma ^2 J \left(12 \eta  J-4 \eta  J e^{-\frac{\kappa  t}{2} }
- \left( 8 \eta  J + 3 \right)e^{\frac{\kappa  t}{4}}
+3\right)^2}{9 \kappa ^2 \left[(4 \eta  J+1) e^{\frac{\kappa  t}{2}}-4 \eta  J\right]}.
\end{equation}
As expected, for no monitoring of the environment ($\eta=0$), one obtains that $\mathcal{Q}[\varrho^{(c)}] \sim J$, i.e. corresponding to the SQL scaling. This function is also monotonically increasing with $t$ and we can expand it around $J=\infty$ to study the leading term, which shows again a quadratic scaling in $J$
\begin{equation}
	\mathcal{Q}[\varrho^{(c)}] \approx \frac{128 \gamma ^2 \eta  J^2 e^{- \kappa t} \left(-3 e^{\frac{\kappa  t}{2}}+2 e^{\frac{3 \kappa  t}{4}}+1\right)^2}{9 \kappa ^2 \left(e^{\frac{\kappa  t}{2}}-1\right)}.
\end{equation}
We also remark that the QFI is equal to the classical FI for a measurement of the quadrature $\hat{P}$, thus showing that a strong measurement of the operator $\hat{J}_z$ on the conditional state of the atomic sample is the optimal measurement saturating the corresponding quantum Cram\'er-Rao bound.
\par
By combining Eqs.~\eqref{eq:fisher_cont} and~\eqref{eq:QFI_t}, we can now define the {\em effective} quantum Fisher information
\begin{align}
\widetilde{\mathcal{Q}} &= \mathcal{F}[p({\bf y}_t)] + \mathbbm{E}_{p({\bf y}_T)}\left[ \mathcal{Q}[\varrho^{(c)}]  \right] = \mathcal{F}[p({\bf y}_t)] + \mathcal{Q}[\varrho^{(c)}], \label{eq:effQFI}
\end{align}
 which represent the inverse of the best achievable variance according to the quantum Cram\'er-Rao bound~\eqref{eq:tcqcrb}.
The resulting expression can be simplified to get the following simple analytical formula
\begin{align}
\label{eq:effQFI_res}
\tilde{\mathcal{Q}} &= K_1 J + \eta K_2 J^2
\end{align}
where
\begin{align}
K_1 &= 32 \frac{\gamma ^2}{\kappa^2} \left(1 - e^{-\kappa t / 4} \right)^2   \,, \label{eq:K1} \\
K_2 &= 64 \frac{\gamma ^2}{\kappa^2} \left(1 - \frac{8}{3} e^{-\kappa t / 4} + 2 e^{- \kappa t / 2} - \frac{1}{3} e^{-\kappa t} \right)  . \label{eq:K2}
\end{align}

\par
\begin{figure}
\centering
\includegraphics[width=8.5cm]{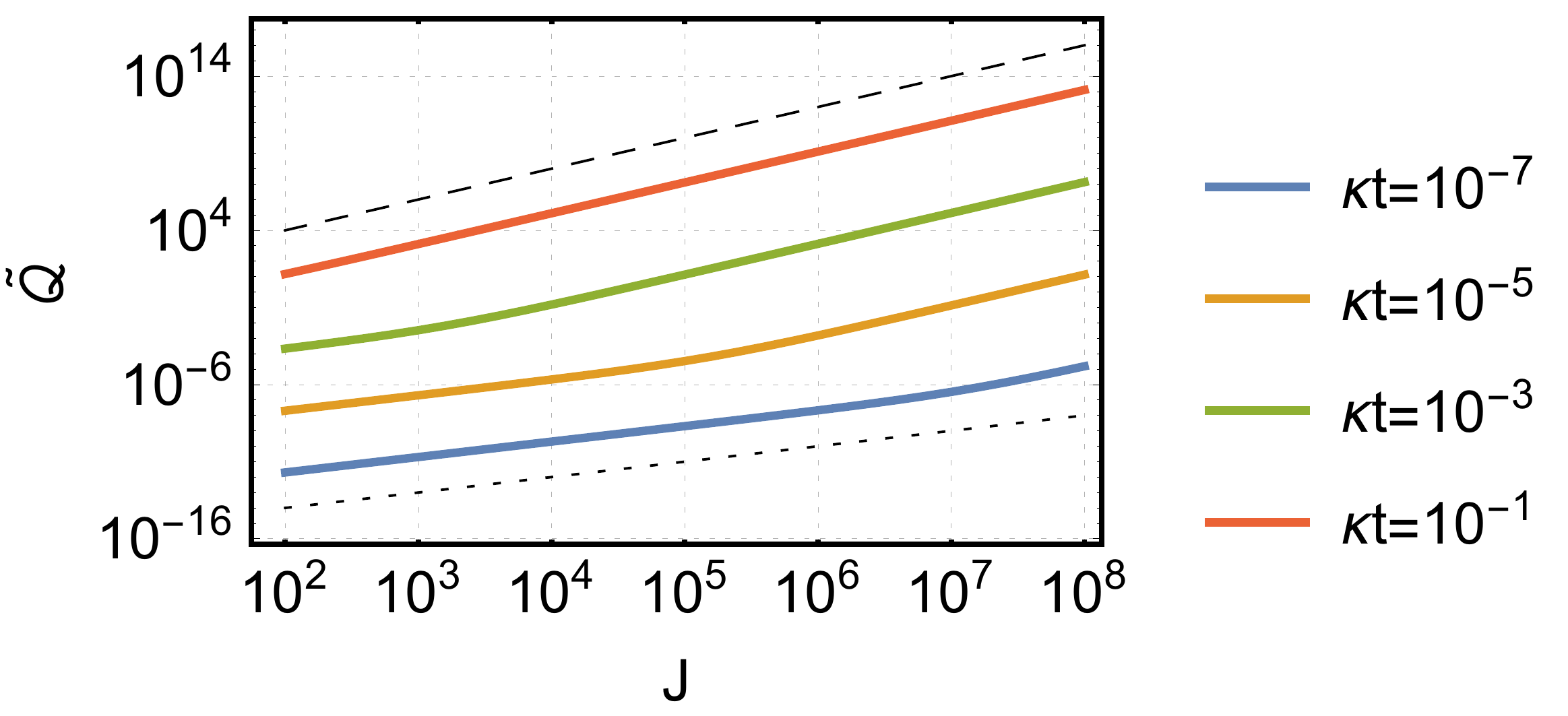}
\caption{{\bf J scaling} - effective QFI $\widetilde{\mathcal{Q}}$ as a function of $J$ for different vales of $\kappa t$, for unit efficiency $\eta$ and effective coupling strength ${\gamma}/{\kappa}=1 \, \mathrm{G}^{-1}$; axes are in logarithmic scale. The solid curves are for increasing values of $\kappa t$ (shown in the legend) from top to bottom.
The two regimes appearing in the plots are $\sim J^2$ (steeper slope) for higher values of $\kappa t$ and higher values of $J$ and $\sim J$ (gentler slope) for the opposite parameters' regions. For visual comparison we show a dashed line at the top $\propto J^2$ and a dotted line at the bottom $\propto J$.}
\label{fig:Jscaling}
\end{figure}

We start by studying how this quantity scales with the total spin: in Fig.~\ref{fig:Jscaling} we plot $\widetilde{\mathcal{Q}}$ as a function of $J$ in the appropriate regions of parameters. We remark that the plots will be presented by using $1/ \kappa$ as a time unit so that the strength of the interaction becomes $\gamma/\kappa$ and is always fixed to $1\, \mathrm{G}^{-1}$ in the following.
We observe that, within the validity of our approximation ($\kappa t \lesssim 1$), it is possible to obtain the Heisenberg-like scaling $J^2$ for the effective QFI. There is a transition between SQL-like scaling and Heisenberg scaling depending on the relationship between $J$ and $\kappa t$ showing how the quantum enhancement is observed for $J \gg 1/\kappa t$.
\par
\begin{figure}
\centering
\includegraphics[width=8.5cm]{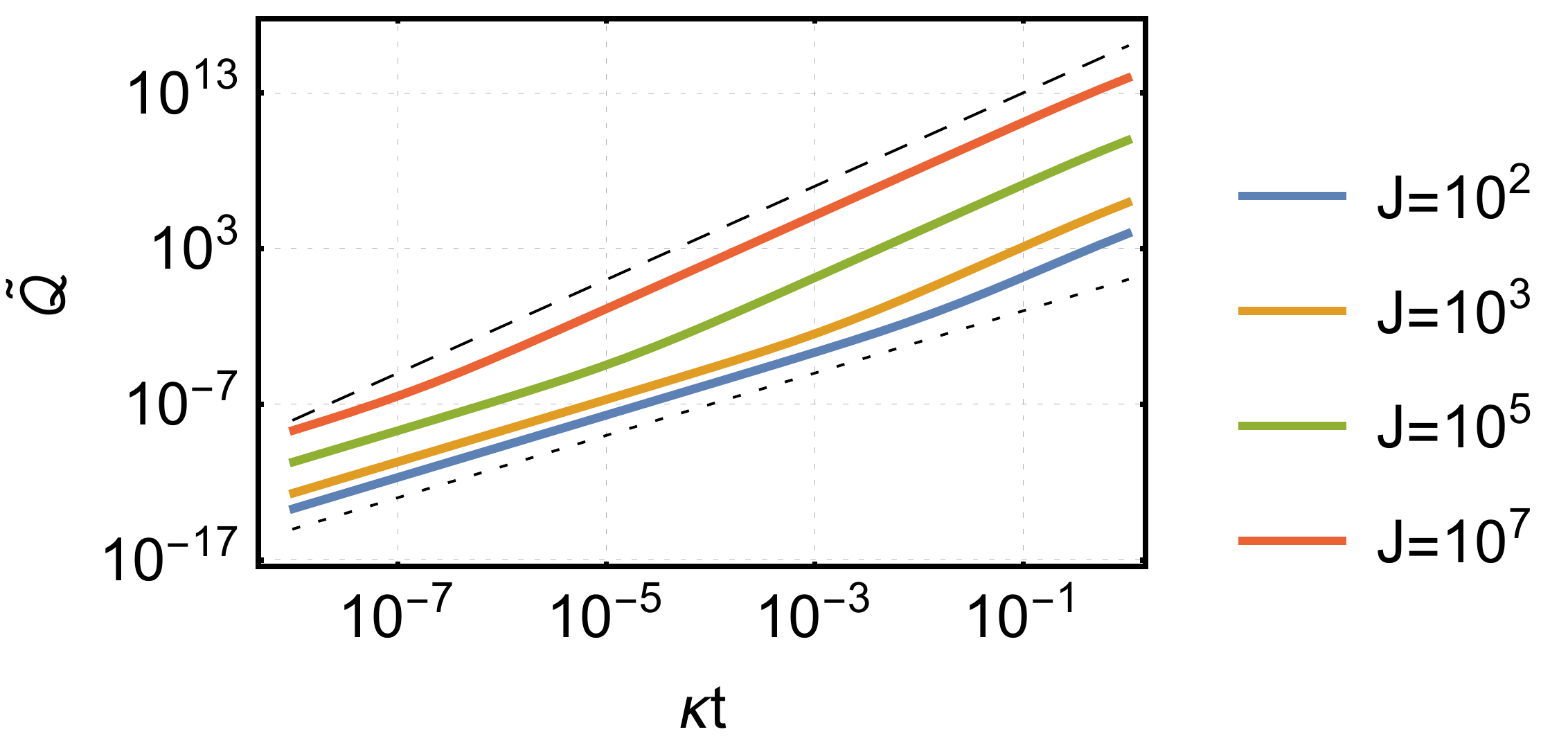}
\caption{{\bf Time scaling} - effective QFI $\widetilde{\mathcal{Q}}$ as a function of $\kappa t$ for different values of $J$, for unit efficiency $\eta$ and effective coupling strength ${\gamma}/{\kappa}=1 \, \mathrm{G}^{-1}$; axes are in logarithmic scale. The solid curves are for increasing values of $J$ (shown in the legend) from top to bottom.
The two regimes appearing in the plots are $\sim (\kappa t)^3$ (steeper slope) for higher values of $\kappa t$ and higher values of $J$ and $\sim (\kappa t)^2$ (gentler slope) for the opposite parameters' regions. For visual comparison we show a dashed line at the top $\propto (\kappa t)^3$ and a dotted line at the bottom $\propto (\kappa t)^2$.}
\label{fig:tscaling}
\end{figure}

The same conclusions are drawn if we look at the behaviour of $\widetilde{\mathcal{Q}}$ as a function of the interrogation time $t$, plotted in Fig.~\ref{fig:tscaling}: a transition from a $t^2$-scaling to a monitoring-enhanced $t^3$-scaling is observed for $J \gg 1/\kappa t$. We remark here that the typical scaling obtained in quantum metrology for unitary parameters is of order $t^2$. The observed $t^3$-scaling is due to the continuos monitoring of the system. A  similar scaling of the Fisher information would be in fact obtained for an equivalent classical estimation problem, where a continuously monitored classical system is estimated via a the Kalman filter \cite{Genoni2017}. Notice that there are also few recent examples in the literature where a $t^4$-scaling can be observed. This is obtained in noiseless quantum metrology problems with time-dependent Hamiltonian and by exploiting open-loop control \cite{Schmitt2017,Pang2017,Yang2017,Gefen2017}. In particular in \cite{Gefen2017}, it was also shown that a $t^3$-scaling can be achieved without additional control, but by performing repeated (stroboscopic) measurement on the system, analogously to our strategy.\\

\par
\begin{figure}[h!]
\centering
\includegraphics[width=8.5cm]{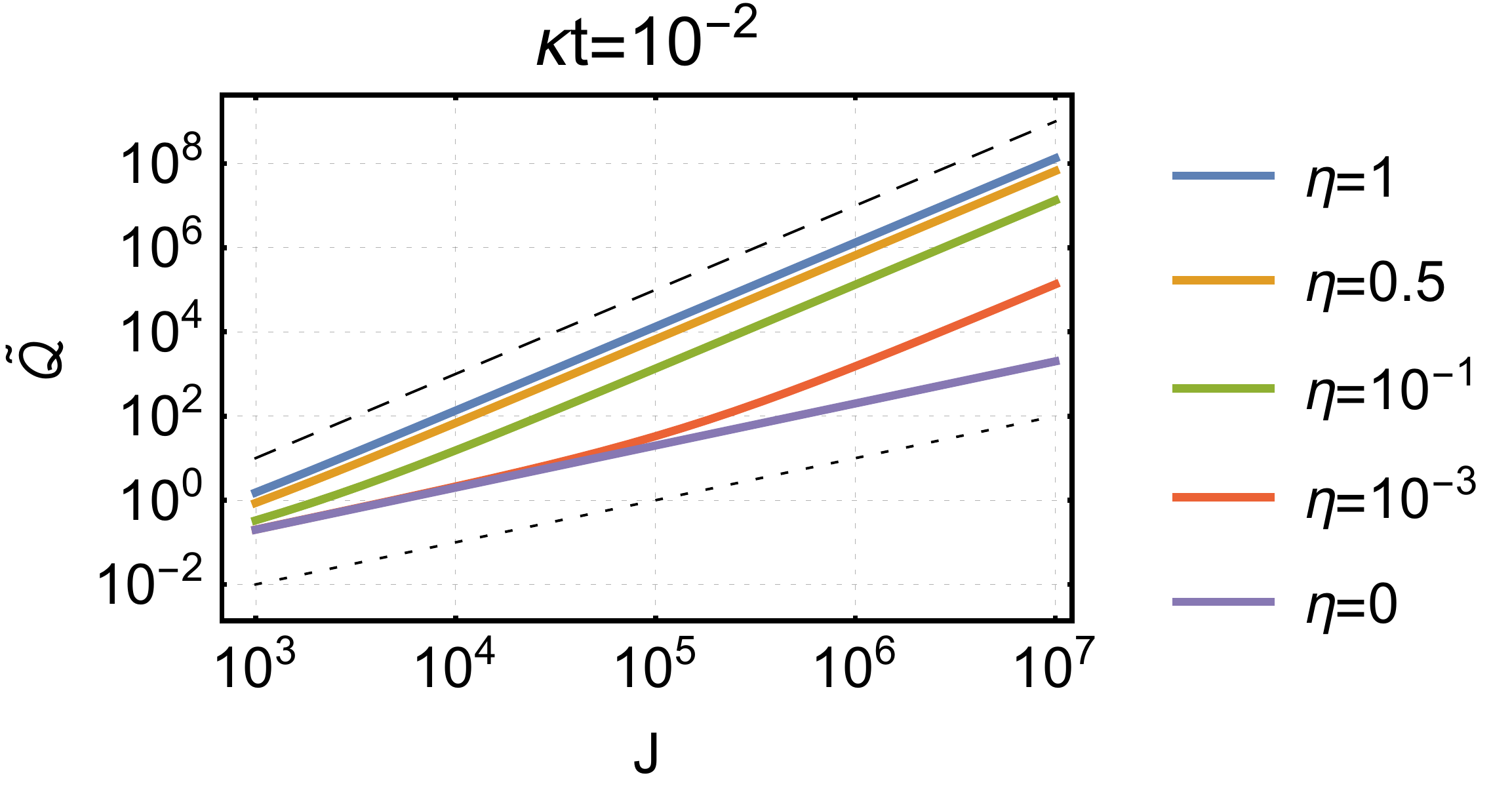}
\includegraphics[width=8.5cm]{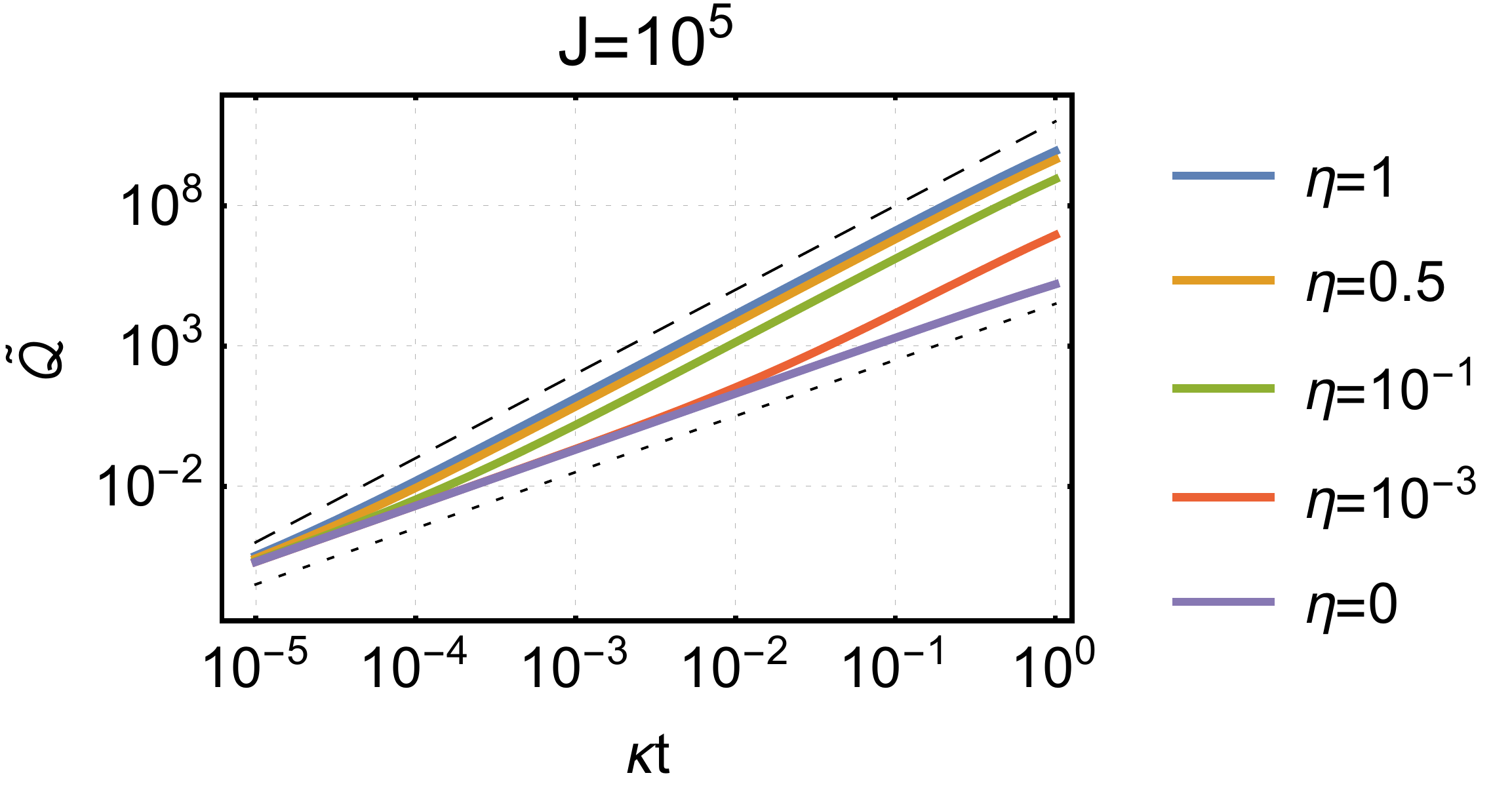}
\caption{{\bf Effect of non unit efficiency} - effective QFI $\widetilde{\mathcal{Q}}$ as a function of $J$ (top panel) and $\kappa t$ (bottom panel) for different values of $\eta$ and effective coupling strength $\gamma/{\kappa}=1 \, \mathrm{G}^{-1}$.
The two regimes appearing in the plots are $\sim J^2$ (top panel) and $\sim (\kappa t)^3$ (bottom panel) for higher values of $\kappa t$ and higher values of $J$ while $\sim J^2$ (top panel) and $\sim (\kappa t)^2$ (bottom panel) for the opposite parameters' regions. For visual comparison we show a dashed line at the top $\propto J^2$ (top panel) and $\propto (\kappa t)^3$ (bottom panel) and also a dotted line at the bottom $\propto J$ (top panel) and $\propto (\kappa t)^2$ (bottom panel).
}
\label{fig:eff}
\end{figure}

The previous results were both shown by considering perfect monitoring of the environment, i.e. for detectors with unit efficiency $\eta$.  In Fig.~\ref{fig:eff} we plot the behaviours of $\widetilde{\mathcal{Q}}$ as a function of $J$ and $t$, varying the detector efficiency $\eta$; we observe how the quantum enhancements can be obtained for all non-zero values of $\eta$. The effect of having a non-unit monitoring efficiency is simply to imply larger values of $J$ to witness the transition between SQL to Heisenberg-scaling, as one can also understand by looking at the role played by $\eta$ and $J$ in Eq. (\ref{eq:effQFI_res}).\\
We remind that if we consider only the classical FI $\mathcal{F}[p({\bf y}_t)]$, the Heisenberg scaling in terms of $J$ and $t^3$-scaling are always obtained for $\kappa t \lesssim 1$ and for every $\eta$, as shown by the expansion~\eqref{eq:fisher_cont_expansionT}.
However, if the contribution of this term, as well as the contribution of conditioning to the QFI, are too small then the QFI of the unconditional state, i.e. \eqref{eq:QFI_t} with $\eta = 0$, dominates (the term $\eta K_2 J^2 $ in \eqref{eq:effQFI_res} is negligible) and we observe SQL scaling for $\mathcal{\tilde{Q}}$.
We finally mention that the in the regimes where we observe Heisenberg scaling of $\mathcal{\tilde{Q}}$, the classical FI $\mathcal{F}[p({\bf y}_t)]$ amounts to a relevant part of the total, namely around 25\% .

\subsection{Optimality of time-continuous measurement strategy for noisy quantum magnetometry}
As explained before, the ultimate limit for quantum magnetometry, in the presence of Markovian transversal noise as the one described by the master equation~\eqref{eq:ME_B}, is given by the QFI $\mathcal{\overline{Q}}_\mathcal{L}$ in Eq.~\eqref{eq:ultimQFI}.
The generalized master equation~\eqref{eq:genME} in this case (considering the large-spin approximation) reads
\begin{equation}
	\label{eq:genMEmagneto}
	\frac{d\bar{\varrho}}{d t} = -i \gamma \sqrt{\bar{J}_t}\left( B_1 \hat{X} \bar{\varrho} - B_2 \bar{\varrho} \hat{X} \right) + \kappa \bar{J}_t \mathcal{D}\left[ \hat{P} \right] \bar{\varrho} \,.
\end{equation}
In~\ref{appendix2} we show how this equation can be solved in a phase space picture, since the equation contains at most quadratic terms in $\hat{X}$ and $\hat{P}$ and thus preserves the Gaussian character of the operator $\bar{\varrho}$.\\ The final result is
\begin{align}
\label{eq:totalQFImag}
\mathcal{\overline{Q}}_{\mathcal{L}_{tn}} &=\mathcal{\tilde{Q}} (\eta=1) = K_1 J + K_2 J^2,
\end{align}
{\em i.e.}, we exactly obtain the effective QFI $\tilde{\mathcal{Q}}$ defined in Eq.~\eqref{eq:effQFI} in the limit of unit efficiency $\eta = 1$.
This result remarkably proves that our strategy, not only allows to obtain the Heisenberg limit, but also corresponds to the optimal one, given a collective transversal noise master equation~\eqref{eq:ME_B} and in the presence of perfectly efficient detectors. Indeed, any other more experimentally complicated strategy, based on entangled and non-local in time measurements of the output modes and the system, would not give better results in the estimation of the magnetic field $B$.
\section{Conclusion and discussion} \label{s:conclusion}
We have addressed in detail estimation strategies
for a static and constant magnetic field acting on an atomic ensemble
of two-level atoms also subject to transverse noise. In particular,
we have evaluated the ultimate quantum limits to
precision for strategies based on time-continuous
monitoring of the light coupled the atomic ensemble.\\
After deriving the appropriate quantum Cram\'er-Rao bound, we
have calculated the corresponding effective quantum Fisher information
in the limit of large spin, posing the ultimate limit on the mean-square
error of any unbiased estimator. Our results conclusively show that both Heisenberg
$J^2$-scaling in terms of spin, and a monitoring-enhanced $t^3$-scaling
in terms of the interrogation time, are obtained for $J \gg 1/\kappa t$, confirming what was discussed in~\cite{Geremia2003,Molmer2004}.
We have remarkably demonstrated that
these quantum enhancements are also obtained  for not unit monitoring efficiency,
i.e. even if one cannot measure all the environmental modes or for not perfectly efficient detectors.
Finally we have analytically proven the
optimality of our strategy, i.e. that given the master equation describing
the unconditional dynamics of the system and ideal detectors, no other measurement strategy
would give better results in estimating the magnetic field.\\
We remark that Heisenberg scaling,
or at least a super-classical scaling, can be obtained
in the presence of collective or individual (independent)
transversal noise, by preparing a highly entangled or spin-squeezed
state at the beginning of the dynamics and, for individual noise,
by optimizing on the interrogation time $t$~\cite{Chaves13,Brask15,Smirne16}.
In this respect, the advantage of our protocol lies in the fact that
it achieves the Heisenberg scaling
even for an initial \emph{classical} spin-coherent state, exploiting
the dynamical spin squeezing that is generated by the weak measurement.
\par
In conclusion, we have shown
that time-continuous measurements represent a resource for noisy
quantum magnetometry~\cite{Geremia2003,Molmer2004,Madsen2004}.
Indeed, the information leaking into the environment, here
represented by light modes coupled to the atomic sample, obtained
via homodyne detection, and the
corresponding measurement back-action on the atomic sample, may
be efficiently (and optimally) exploited
in order to obtain the promised quantum enhanced estimation precision.
\section*{Acknowledgments}
MGG would like to thank A. Doherty and A. Serafini for discussions
and acknowledges support from Marie Sk\l odowska-Curie Action
H2020-MSCA-IF-2015 (project ConAQuMe, grant nr. 701154).
This work has been supported by EU through the collaborative H2020
project QuProCS (Grant Agreement 641277) and by UniMI through the H2020 Transition Grant.

\appendix
\section{Classical and quantum Cram\'er-Rao bounds for sequential non-demolition measurements}
\label{appendix0}
Here we will show how to derive the quantum Cram\'er-Rao bound
for time-continuous homodyne monitoring  reported in Eq.~\eqref{eq:tcqcrb}.\\ 
We start by considering a (classical) estimation problem of a parameter
$\theta$ described by a conditional probability $p(z ,{\bf y}_T
| \theta)$, where the vector ${\bf y}_T = (y_1,y_2,\dots,y_T)^{\sf T}$
contains the outcomes of sequential measurements performed up to time
$T$, while $z$ corresponds to a final measurement performed on the state
of the system that has been conditioned on the previous measurement
results ${\bf y}_T$. The corresponding classical Fisher information
can be evaluated as
\begin{align}
\mathcal{F}[p&(z,{\bf y}_T| \theta)] = \int d{\bf y} \,dz\, p( z, {\bf y}_T | \theta) \left( \partial_\theta \log p( z, {\bf y}_T | \theta) \right)^2  \nonumber\\
&= \int d{\bf y} \,dz \, p( z | {\bf y}_T , \theta) p({\bf y}_T | \theta) \left[ \left(  \partial_\theta \log p( z | {\bf y}_T, \theta) \right)^2 \right.  \nonumber \\
& \:\:\;\;\:\: + 2  ( \partial_\theta \log p( z | {\bf y}_T, \theta) )\, (\partial_\theta \log p({\bf y}_T | \theta)) \nonumber \\
& \:\:\;\;\:\: + \left. \left(  \partial_\theta \log p({\bf y}_T | \theta)  \right)^2 \right] \,
\end{align}
where the second expression has been obtained by means of the Bayes rule
$$p( z, {\bf y}_T | \theta) = p( z | {\bf y}_T , \theta) p({\bf y}_T | \theta)\,.$$

In the following, we will omit the dependence on the parameter $\theta$ and
we will denote
by $\mathbbm{E}_{p(x)}[\cdot]$ the average over a probability distribution $p(x)$.
By considering each term inside the integral separately one obtains

\begin{align}
 &\mathbbm{E}_{p(z,{\bf y}_T)} \left[ \left( \partial_\theta \log p( z | {\bf y}_T) \right)^2\right] =
\mathbbm{E}_{p({\bf y}_T )} \left[ \mathcal{F}\left[p(z | {\bf y}_T) \right]  \right] \,
\\
&2 \,\mathbbm{E}_{p(z,{\bf y}_T)}  \left[ \partial_\theta \log p( z | {\bf y}_T) \, \partial_\theta \log p({\bf y}_T ) \right] = \\
& = 2 \,\int d{\bf y} \, \left(\partial_\theta p({\bf y}_T) \right) \int dz \, \left(\partial_\theta p( z | {\bf y}_T )\right)  =0 \,
\nonumber \\ 
 & \mathbbm{E}_{p(z,{\bf y}_T)}  \left[ \left( \partial_\theta \log p({\bf y}_T) \right)^2\right] = \mathcal{F}[p({\bf y}_T)]  \,
\end{align}

where we have used the property $ \int dz \, \left(\partial_\theta p( z |
{\bf y}_T )\right) = \partial_\theta \int dz \, p( z | {\bf y}_T ) =
\partial_\theta \,(1) = 0$. As a consequence, any unbiased estimator
$\hat{\theta}$ based on $M$ experiments, i.e. obtained collecting $M$ series
of measurement outcomes $({\bf y}_T, z)$, satisfies the generalized Cram\'er-Rao
bound
\begin{align}\label{gcr}
{\rm Var}_{\hat{\theta}}(\theta) \geq \frac{1}{M\left( \mathcal{F}[p({\bf y}_T)] + \mathbbm{E}_{p({\bf y}_T)}\left[ \mathcal{F}[p(z | {\bf y}_T)]  \right]\right) }\,
\end{align}
where the first term $\mathcal{F}[p({\bf y}_T)]$ is the Fisher information
corresponding to the sequential measurements with outcomes ${\bf y}_T$, while
the second term is the average of the Fisher information
$\mathcal{F}[p(z | {\bf y}_T)]$, corresponding to the final measurement
over all the possible trajectories conditioned on the previous measurement
results ${\bf y}_T$. The bound in Eq.~\eqref{gcr} bears some formal similarity
to the Van Tree's inequality~\cite{VanTree}, which however applies in a quite
different situation, i.e. the case where the parameter to be estimated
$\theta$ is a random variable distributed according to a given probability
distribution $p(\theta)$. \\

\par
The estimation strategy here described is of particular interest when we deal with quantum systems, given the back-action of quantum measurement on the state of the system itself. We can in fact associate each measurement outcome ${y}_k$ to a Kraus operator $M_{y_k}$ such that the conditional quantum state, for the system initially prepared in a state $\varrho_0$ and after obtaining the stream of outcomes ${\bf y}_T$, reads
\begin{align}
\varrho^{(c)}_{{\bf y}_T}  = \frac{\tilde{M}_{{\bf y}_T} \varrho_0 \tilde{M}_{{\bf y}_T} ^\dag}{\Tr[\tilde{M}_{{\bf y}_T} \varrho_0 \tilde{M}_{{\bf y}_T} ^\dag]}
 \:. \label{eq:discrete}
\end{align}
where $\tilde{M}_{{\bf y}_T} = M_{y_T} \dots M_{y_2} M_{y_1}$ and the probability of obtaining the outcomes ${\bf y}_T$ reads $p({\bf y}_T | \theta) =\Tr[\tilde{M}_{{\bf y}_T} \varrho_0 \tilde{M}_{{\bf y}_T} ^\dag]$ \footnote{In our treatment we will consider the sequential non-demolition measurement fixed, and thus described by a fixed set of Kraus operators $\{M_{y_k}\}$. However one can generalize the results by considering adaptive schemes where one can decide to modify the measurement performed at each time $t_k$}.
One can then also perform a {\em strong} (destructive) measurement described by POVM operators $\{\Pi_z\}$ on the conditional state, and the whole measurement strategy is described by the conditional probabilities
\begin{align}
p(z | {\bf y}_T,  \theta) &= \Tr[ \varrho^{(c)}_{{\bf y}_T} \Pi_z ] \,, \nonumber \\
p(z , {\bf y}_T | \theta) &= p(z | {\bf y}_T  ,\theta) \,p({\bf y}_T | \theta) \nonumber \\
&= \Tr[\tilde{M}_{{\bf y}_T} \varrho_0 \tilde{M}_{{\bf y}_T} ^\dag \Pi_z]
\end{align}
Typically the parameter to be estimated $\theta$ enters in the the dynamics described by the Kraus operators $M_{y_k}$. For this reason we will start by considering these operators fixed, while we suppose we can optimize over the final measurement $\{\Pi_z\}$. We can then apply the quantum Cramér-Rao bound for the conditional states $\varrho^{(c)}_{{\bf y}_T}$, stating that $\mathcal{F}[p(z | {\bf y}_T)]  \leq \mathcal{Q}[\varrho^{(c)}_{{\bf y}_T} ]$. One then obtains a more fundamental quantum Cram\'er-Rao bound for our estimation strategy
\begin{align}
{\rm Var}_{\hat{\theta}}(\theta) \geq \frac{1}{M\left( \mathcal{F}[p({\bf y}_T)] + \mathbbm{E}_{p({\bf y}_T)}\left[ \mathcal{Q}[\varrho^{(c)}_{{\bf y}_T}]  \right]\right) }\, .
\label{eq:tcqcrbAPP}
\end{align}
Clearly this bound can be readily applied to the time-continuous case discussed in the main text, where the vector of outcomes ${\bf y}_T$ corresponds to a measured homodyne photocurrent, and where the conditional state $\varrho^{(c)}_{{\bf y}_T}$ can be obtained via a stochastic master equation as the one in Eq.~\eqref{eq:SME}.\\
We should also remark that a bound of this kind has already been considered in~\cite{Catana2014}, in a similar physical situation where $n$ probes, that may be prepared in a quantum correlated initial state, are coupled to $n$ independent environments and one performs sequentially
$n$ measurement on the respective environments and a final measurement on the conditional state of the probes.
\section{Gaussian dynamics and Gaussian Fisher information}
\label{appendix1}
Here we will provide the formulas for the dynamics of the atomic ensemble described by the stochastic master equation~\eqref{eq:stochME_spin}.
As we mentioned in the text, the whole dynamics preserves the Gaussian character of the quantum state and thus can be fully described in terms of the first moments vector $\langle {\bf \hat{r}}\rangle_c$ and of the covariance matrix $\sigmaCM$ of the quantum state $\varrho^{(c)}$. These are defined in components as $\langle \hat{r}_j\rangle_c={\rm Tr}\left[\hat{r}_j \varrho^{(c)} \right]$ and $\sigma_{jk}={\rm Tr}\left[\{\hat{r}_j - \langle \hat{r}_j \rangle_c ,\hat{r}_k - \langle \hat{r}_k \rangle_c\}\varrho^{(c)}\right]$ for the operator vector $\hat{\bf r}=(\hat{X},\hat{P})^{\sf T}$.
In formulae one obtains \cite{diffusone,WisemanDoherty}:
\begin{align}
d\langle {\bf \hat{r}}\rangle_c &= {\bf u}\, dt + \frac{\sigmaCM M \, d{\bf w}}{\sqrt{2}} \:,  \\
\frac{d \sigmaCM}{dt} &=  D - \sigmaCM M M^{\sf T} \sigmaCM    \:,  \label{eq:evolutionCM}
\end{align}
where
\begin{align}
D &=
\left(
\begin{array}{c c}
2 \kappa J e^{-\kappa t/2} & 0 \\
0 & 0 \\
\end{array}
\right) , \\
M &=
\left(
\begin{array}{c c}
0 & 0 \\
\sqrt{2 \eta \kappa J e^{-\kappa t/2}} & 0
\end{array}
\right), \\
{\bf u} &= ( 0, -\gamma B \sqrt{J e^{-\kappa t/2}})^{\sf T} \,,
\end{align}
and $d{\bf w}$ is a vector of Wiener increments such that $d{w}_j \,dw_k =\delta_{jk} dt$, related to the photocurrent via the equation
\begin{align}
d{\bf y}_t = - M^{\sf T} \langle {\bf \hat{r}}\rangle_c \,dt + \frac{d{\bf w}}{\sqrt{2}} \,.
\end{align}
The Eqs.~\eqref{eq:firstmom}, \eqref{eq:secondmom} and \eqref{eq:derR} 
can be obtained from the ones above, remembering that for our definitions $\sigma_{22} = 2 {\rm Var}_c[\hat{P}(t)]$.\\

The method to calculate the Fisher information corresponding to the time-continuous measurement in the case of linear Gaussian system has been described in \cite{Genoni2017}. One has to evaluate the formula
\begin{align}
\mathcal{F}[p({\bf y}_t)] = \mathbbm{E}_{p({\bf y}_t)} \left[ 2 (\partial_B \langle {\bf \hat{r}}\rangle_c^{\sf T} ) M M^{\sf T} ( \partial_B \langle {\bf \hat{r}}\rangle_c) \right] \,,
\end{align}
that, by plugging in the matrices describing our problem, is easily simplified to Eq.~\eqref{eq:tcfisher_magenot}. \\
As the conditional state is Gaussian, also the calculation of the corresponding QFI can be easily obtained, in this case by applying the results presented in \cite{Pinel2013}. Moreover, as only the first moments of the state depend on the parameter $B$, the calculation is further simplified and one has
\begin{align}
\mathcal{Q}[\varrho^{(c)}] = 2\,  (\partial_B \langle {\bf \hat{r}}\rangle_c^{\sf T} ) \, \sigmaCM^{-1} \, (\partial_B \langle {\bf \hat{r}}\rangle_c) \,.
\end{align}
By noticing that the only non-zero entry of the vector $\partial_B \langle {\bf \hat{r}}\rangle_c$ is the one corresponding to $\langle \hat{P}(t) \rangle_c$, one easily obtain Eq.~\eqref{eq:QFI_formula}.
\section{Bayesian analysis for continuously monitored quantum systems}\label{appendixBayes}
Bayesian analysis has proven to be an efficient tool for estimation in continuously monitored quantum systems~\cite{Chase2009,GammelmarkCRB,KiilerichHomodyne,Cortez2017}.
The goal is to reconstruct the posterior distribution of $B$ given the observed current $\mathbf{y}_t$, by Bayes rule:
\begin{equation}
\label{eq:bayes}
p \left( B | \mathbf{y}_t  \right) = \frac{L( B | \mathbf{y}_t) p (B)}{p(\mathbf{y}_t)},
\end{equation}
where $p( B)$ is the prior distribution, $L( B | \mathbf{y}_t) \equiv p( \mathbf{y}_t | B)$ is the likelihood and $p(\mathbf{y}_t)$ serves as a normalization factor.
The Bayesian estimator is the mean of the posterior distribution $\hat{B}( \mathbf{y}_t ) 	= \mathbbm{E}_{p \left( B | \mathbf{y}_t  \right)} \left[ B \right] $ and it is proven that the corresponding variance $\mathrm{Var}_{\hat{B} }(B)= \mathbbm{E}_{p \left( B | \mathbf{y}_t  \right)} [ B^2 ] - (\mathbbm{E}_{p \left( B | \mathbf{y}_t  \right)} [ B])^2$ is asymptotically optimal, {\em i.e.} tends to saturate the Cram\'er-Rao bound when the length of the vector ${\bf y}_T$ is large.
\par
The simulated experimental run is obtained by numerically integrating the stochastic differential equation \eqref{eq:firstmom} with the Euler-Maruyama method for the ``true'' value of the parameter $B_\text{true}$. Time is discretized with steps of length $\Delta t$, i.e. to get from time $0$ to time $T$ we perform $n_{T} = T/{\Delta t}$ steps. Experimental data is represented by the observed measurement current $\mathbf{y}_T = (\Delta y_{t_{1}},\dots,\Delta y_{t_{n_T}} )^{\sf T}$, which corresponds to an $n_{T}$-dimensional vector.
The outcome at every time step $\Delta y_{t_{i}}$ is sampled from a Gaussian distribution with variance $\Delta t$ and mean $\overline{\Delta y_{t_i}}(B) = \sqrt{2 \eta \kappa J e^{-\frac{\kappa t}{2}}} \langle \hat{P}(t_{i}) \rangle_c \Delta t$. Notice that $\overline{\Delta y_{t_i}}(B)$ depends explicitly on the parameter $B$ via the quantum expectation value  $\langle \hat{P}(t_{i}) \rangle_c$ on the conditional state.
\par
Since we are estimating only one parameter the posterior can be obtained on a grid on the parameter space, while for more complicated problems Markov chain Monte Carlo methods might be needed to sample from the posterior \cite{GammelmarkCRB}. In practical terms we need to solve Eqs.~\eqref{eq:firstmom} and \eqref{eq:secondmom} for every value of the parameter $B$ on the grid, assuming to perfectly know all the other parameters; then we need to calculate the likelihood for each value via
\begin{equation}
\label{eq:lklhood}
L\left( B | \mathbf{y}_T \right) \propto \prod_{i=0}^{n_{T}} \exp \left[ - \frac{\left(\Delta y_{t_i} - \overline{\Delta y_{t_i}}(B) \right)^2  }{2 \,\Delta t} \right],
\end{equation}
by considering the outcomes as independent random variables, i.e. multiplying the corresponding probabilities.
We then apply Bayes rule, Eq.~\eqref{eq:bayes}, assuming a flat prior distribution $p(B)$ on a finite interval. The same analysis is trivially applied to more than one experiment by simply multiplying the likelihood obtained for every different observed measurement current.
\section{Ultimate quantum Fisher information via generalized master equation in phase space
\label{appendix2}
}
Here we explicitly show how to solve Eq.~\eqref{eq:genMEmagneto}.
The characteristic function for a generic operator $\hat{O}$ is defined as
\begin{equation}
\label{eq:charfundef}
	\chi[\hat{O}](\mathbf{s}) = \Tr \left[ \hat{D}_{-\mathbf{s}} \hat{O}\right],
\end{equation}
where the displacement operator is defined as
\begin{equation}
\hat{D}_{-\mathbf{s}}=\exp\left( i \mathbf{s}^\top \Omega \hat{\mathbf{r}} \right).
\end{equation}
In particular we will work in the phase space of a single mode system, so that $\hat{\mathbf{r}}^\top = (\hat{X},\hat{P})$ is the vector of quadrature operators and $\mathbf{s}^\top=(x,p)$ is the vector of phase space coordinates.\\
The action of operators in the Hilbert space corresponds to differential operators acting on the characteristic function via the following mapping~\cite{diffusone,Barnett1997}
\begin{align}
\hat{X} \rho &\leftrightarrow \left(- i \partial_p - \frac{x}{2} \right) \chi(\mathbf{s}) \\
\rho \hat{X} &\leftrightarrow \left(- i \partial_p + \frac{x}{2} \right) \chi(\mathbf{s})\\
\hat{P} \rho &\leftrightarrow \left( i \partial_x  - \frac{p}{2} \right) \chi(\mathbf{s})\\
\rho \hat{P} &\leftrightarrow \left(i \partial_x  + \frac{p}{2} \right) \chi(\mathbf{s}).
\end{align}

If we now define the characteristic function associated to the operator $\bar{\varrho}$ introduced in Eq.~\eqref{eq:genME}
\begin{equation}
	\bar{\chi} \left( \mathbf{s},t \right) \equiv \chi \left[ \bar{\varrho} \right](\mathbf{s}),
\end{equation}
the quantity of interest in order to compute the QFI is then $\Tr \bar{\varrho} = \bar{\chi}(0,t)$, as evident from the definition \eqref{eq:charfundef}.

By applying the phase space mapping, from the generalized master equation \eqref{eq:genMEmagneto} we get to the following partial differential equation for the characteristic function
\begin{align}
&\frac{d\bar{\chi}(\mathbf{s},t)}{dt} =  \nonumber \\
& =\left[i \gamma \sqrt{\bar{J}_t} \frac{B_1 + B_2}{2} x - \frac{\kappa \bar{J}_t}{2} p^2 - \gamma \sqrt{\bar{J}_t} \left( B_1 - B_2\right) \partial_p \right] \bar{\chi}(\mathbf{s},t). \label{eq:ChF_MEfinal}
\end{align}
This equation can be solved by performing a Gaussian ansatz, similarly to \cite{Guarnieri2016}, i.e. assuming that at every time the characteristic function can be written in the following form
\begin{align}
\label{eq:gauss_ansatz}
\bar{\chi}(\mathbf{s},t)=C(t) \exp\Biggl[ - \frac{1}{4} \mathbf{s}^{\top} \Omega^\top  \sigmaCM(t) \Omega \, \mathbf{s} \,  + \nonumber \\
+ \, i \mathbf{s}^{\top} \Omega^\top \mathbf{s}_m(t)\Biggr].
\end{align}
The dependence on time and on the parameters $B_{1/2}$ is completely contained in the covariance matrix $\sigmaCM(t)$, in the first moment vector $\mathbf{s}_m(t)^\top = \left(x_m(t),p_m(t)\right)$ and in the function $C(t)=\bar{\chi}(0,t)$, which is the final result we are seeking.

By plugging \eqref{eq:gauss_ansatz} into \eqref{eq:ChF_MEfinal} and equating the coefficients for different powers of $x$ and $p$, one obtains a system of differential equations; the relevant ones are the equations coming from the coefficients of order one, and multiplying $p$ and $p^2$:
\begin{align}
& \dot{\mathbf{\sigma}}_{1,1}(t)=2 \kappa J e^{- \frac{\kappa t }{2}} \\
& \dot{x}_m(t) = - i \frac{\gamma}{2} \sqrt{J e^{- \frac{\kappa t }{2}}} \left( B_1 - B_2\right) \mathbf{\sigma}_{1,1}(t)\\
& \dot{C}(t) = - i  \gamma \sqrt{J e^{- \frac{\kappa t }{2}}} \left(B_1 - B_2\right) x_m(t) C(t) \,.
\end{align}
These equations are solved analytically with the initial conditions $\mathbf{\sigma}_{1,1}(0)=1$, $x_m(0)=0$ and $C(0)=1$ (since for $t=0$ the operator $\bar{\varrho}$ corresponds to the initial state of the system $|0 \rangle \langle 0 |$), yielding
\begin{align}
C(t) = \exp \biggl[-\frac{4 \gamma ^2}{3 \kappa ^2} J (B_1-B_2)^2 e^{- \kappa t} \left(e^{\frac{\kappa  t}{4}}-1\right)^2 \cdot \nonumber \\
\cdot \left(-4 J e^{\frac{\kappa  t}{4}}+(6 J+3) e^{\frac{\kappa  t}{2}}-2 J\right)\biggr]\,.
\end{align}
By plugging this term into Eq.~\eqref{eq:ultimQFI}, we finally obtain the ultimate QFI $\overline{\mathcal{Q}}_{\mathcal{L}_{tn}}$ reported in Eq.~\eqref{eq:totalQFImag}.

\bibliography{LSMbiblio}

\end{document}